\documentstyle[aps,preprint,pra,epsfig]{revtex}

\begin{document}

\title{Topological Filters for Solitons in Coupled Waveguides 
Networks}

\author{R. Burioni, D. Cassi, A. Trombettoni, and A. Vezzani}

\address{I.N.F.M. and Dipartimento di Fisica, Universit\`a
di Parma, parco Area delle Scienze 7A Parma, I-43100, Italy}

\author{P. Sodano}

\address{Dipartimento di Fisica and Sezione I.N.F.N.,
Universit\`a di Perugia, Via A. Pascoli Perugia, I-06123, Italy}

\maketitle

\begin{abstract}
We study the propagation of discrete solitons on chains of coupled 
optical waveguides where finite networks of waveguides are inserted 
at some points. 
By properly selecting the topology of 
these networks, it is possible to control the transmission
of traveling solitons: we show here that 
inhomogeneous waveguide networks may be 
used as filters for soliton propagation. Our results provide 
a first step in the understanding of the interplay/competition between 
topology and nonlinearity for soliton dynamics in optical fibers.
\end{abstract} 

The discovery of solitons 
in optical fibers, three decades 
ago \cite{hasegawa73}, stimulated a huge amount of work aimed at 
using solitons for high speed 
communications \cite{haus96}.  Many 
experiments evidenced the role of the Kerr nonlinearity in allowing for 
the propagation over long-distances of solitons in optical fibers 
\cite{haus96,kivshar03}; however, this nonlinear paradigm 
has not yet demonstrated decisively its advantages over other  
more conventional signal propagation schemes. 
If one introduces a spatial inhomogeneity 
in the field equations by the linear compression mechanism, 
one could stabilize the soliton 
propagation \cite{hasegawa98}.
Motivated by this, one is lead to investigate the so-called 
{\it dispersion-managed} nonlinear Schr\"odinger equation \cite{ablowitz98}
\begin{equation}
i \frac{\partial E}{\partial z}=-\frac{\beta(z)}{2} 
\frac{\partial^2 E}{\partial t^2} - \nu \mid E \mid^2 E,
\label{NLSE}
\end{equation}
where $E(z,t)$ is the electric field, $\nu$  
the Kerr nonlinearity, $z$ the propagation 
direction and $t$ the retarded time. $\beta(z)$ is the 
term responsible for dispersion management. 
In Eq.(\ref{NLSE}) the term $\beta(z)$ plays the role  
of a spatial modulation of the kinetic term; upon discretization, 
this naturally 
leads to consider a network arranged in a non-translational 
invariant topology. 

In this paper we shall consider a discrete version of the nonlinear 
Schr\"odinger equation (\ref{NLSE}), describing an array 
of coupled optical waveguides \cite{hennig99}. Namely,  
we shall show that, by choosing 
a pertinent network of coupled waveguides, one may engineer soliton filters 
for discrete soliton propagation. 
We focus on waveguide networks, built by adding 
a finite discrete network of optical waveguides (the graph $G^0$) 
to a chain (see Fig.1).
By means of a a general criterion derived within a linear approximation 
for a relevant class of solitons \cite{burioni05}, 
denoted as {\it large-fast solitons},
we show the possibility of controlling 
the soliton scattering, by suitably choosing the topology of $G^0$. 
The analytical findings are in agreement with the numerical study 
of the full nonlinear evolution equation. 

The discrete nonlinear Schr\"odinger equation (DNLSE), for a general 
network of coupled optical waveguides, reads  
\begin{equation} 
\label{DNLS-gen} 
i \frac{\partial E_n}{\partial z} = - \sum_{j}  \beta_{n,j}
E_{j}+ \Lambda \mid E_n \mid ^2 E_n.
\end{equation}   
Here $E_n(z)$ is the electric field in the $n$th waveguide 
and $\Lambda$ is proportional to the Kerr nonlinearity. 
The normalization is chosen to be $\sum_n \mid E_n(z) \mid^2=1$.  
In Eq.(\ref{DNLS-gen}), $\beta_{n,j}$ is proportional to the 
mode overlap of the electric fields of the waveguides $n$ and $j$ 
\cite{hennig99,ablowitz03} and it is non-$0$ only
if $n$ and $j$ are neighbours waveguides.
If identical waveguides are arranged to form a chain, then 
Eq.(\ref{DNLS-gen}) assumes the usual form 
$i \partial E_n / \partial z = - \beta_c (E_{n+1}+E_{n-1}) 
+ \Lambda \mid E_n \mid ^2 E_n$ 
where $\beta_{n,n \pm 1}=\beta_c$. 
If, on the other hand, the waveguides are arranged on the sites of a 
non-translational invariant network, a space modulation of the kinetic
term occurs, even if $\beta$ does not depend on 
$z$. As an example, one can consider the waveguide 
geometry depicted in Fig.1. Of course, one can imagine and engineer 
a huge variety of network topologies; our aim here is to show 
that the network topology may be used to engineer a novel class of filters 
for soliton motion over spatially inhomogeneous networks of coupled waveguides.
 
It is well known that the DNLSE on a homogeneous chain 
is not integrable \cite{hennig99}; nevertheless,
soliton-like wavepackets can propagate for a long time 
\cite{malomed96}, and the stability conditions of these 
soliton-like solutions may be analyzed
within a standard variational approach \cite{malomed96,trombettoni01}. 
Let us prepare, at $z=0$, a Gaussian wavepacket: 
$E_{n}(z=0) \propto 
\exp{ \{ -\frac{(n-\xi_0)^2}{\gamma_0^2} + ik_0(n-\xi_0) \}}$.
If $\Lambda>0$ ($\Lambda<0$), one has a soliton solution 
only if $\cos{k_0}<0$ ($\cos{k_0}>0$). In the following, we assume 
$\Lambda>0$ and $\pi/2 \le k_0 \le \pi$ (positive velocities). 
For $\gamma_0 \gg 1$, the variational soliton-like solution 
should satisfy the relation: 
\begin{equation}
\label{lambda_sol}
\Lambda_{sol} \approx  \beta_{c} \sqrt{\pi} \frac{\mid \cos{k_0} \mid} 
{\gamma_0}.
\end{equation}
The long-time stability of variational solutions has been  
numerically checked. Hereafter, the term ``solitons'' denotes such 
variational solutions. Interestingly, 
discrete solitons have been experimentally 
observed in optical waveguides \cite{eisenberg98}. 

Let us consider the inhomogeneous network 
obtained by attaching a finite graph $G^0$ to 
a single site of a chain (see Fig.1). 
We denote the generic waveguide with latin indices $i,j,\dots$. 
The index $i$ can be an integer number or a greek letter
$\alpha, \beta, \dots$ according to if the waveguide $i$ belongs to the 
chain or to $G^0$ respectively.
A single link connects the waveguides $0$ 
of the chain with the waveguide $\alpha$ of $G^0$. 
We suppose the waveguides of the infinite chain 
to be identical, so that their coupling terms are set to the 
constant value $\beta_c$. 
Soliton propagation in a generic inhomogeneous network 
is conveniently described within graph theory \cite{harary69}. 
The adjacency matrix ${ A }^0$ 
of $G^0$ is defined as: ${ A }^0_{i,j}=\beta_{i,j}$ 
when $i$ and $j$ are nearest-neighbours sites of $G^0$,
and $0$ otherwise. $G^r$ denotes the graph obtained by 
cutting the site $\alpha$ from $G^0$, and $A^r$ is 
its adjacency matrix (see Fig 1).
The energy levels of $G^0$ and $G^r$ are 
the eigenvalues of $A^0$ and  $A^r$ respectively.

The scattering of a soliton through this topological 
perturbation has been numerically
studied in the following way. At $z=0$, 
we prepare a Gaussian soliton, 
far left from $0$ (i.e., $\xi_0<0$), moving
towards $n=0$ ($\sin(k_0)>0$), and with a width 
$\gamma_0$ related to the nonlinear coefficient $\Lambda$ through 
Eq.(\ref{lambda_sol}). 
We numerically evaluate the nonlinear evolution of the electric field 
from Eq.(\ref{DNLS-gen}). 
The group velocity is $v=\beta_c \sin{k_0}$ and, when 
$z_s \approx |\xi_0| /\sin(k_0)$, the soliton scatters through the 
finite graph $G^0$. At a position $z$ well after
the soliton scattering ($z \gg z_s$), we evaluate
the reflection and transmission coefficients 
${\cal R}=\sum_{n<0} \mid E_n(z) \mid^2$
and ${\cal T}=\sum_{n>0} \mid E_n(z) \mid^2$. 

The interaction between the soliton 
and the defect is characterized by two length scales 
\cite{miroshnichenko03}:  
the soliton-defect interaction length
$z_{int}=\gamma_0/ \beta_c \sin{k}$ and the soliton dispersion 
length $z_{disp}= \gamma_0/ (4 \beta_c \sin{(1/2\gamma_0) \cos{k})}$. 
When $\gamma_0 \gg 1$ and $z_{int} \ll z_{disp}$, the soliton is a 
large-fast soliton: i.e, during the scattering it may be regarded 
as a set of non interacting plane 
waves and the transmission coefficients may be computed 
by considering the transport of a plane wave across the 
topological defect. The results obtained within the linear approximation are 
in good agreement with the numerical solution 
of Eq.(\ref{DNLS-gen}) [see Figs.3-4]. 
We point out that, even if a linear approximation is used \cite{burioni05}, 
the nonlinearity still plays a role: it keeps together the soliton  
during its propagation (see Fig.2). 

In the large-fast soliton regime, the momenta 
for perfect reflection and transmission are completely determined 
by the spectral properties of the graph $G^0$ \cite{burioni05}. 
The linear eigenvalue equation to investigate is  
$- \sum_{m} A_{n,m} E_m=\mu E_n $, where $A_{i,j}=\beta_{i,j}$ 
is the generalized adjacency matrix of the whole network. The momenta 
$k$ corresponding to perfect reflection (${\cal R}(k)=1$) 
and to perfect transmission (${\cal T}(k)=1$) of a plane wave 
are determined by imposing the continuity 
at sites $0$ and $\alpha$. One obtains ${\cal R}=1$ if 
$2 \beta_c \cos{k}$ coincides with an energy level of $G^0$, while
${\cal T}=1$ if  $2 \beta_c \cos{k}$ is an energy level 
of the reduced graph $G^r$ \cite{burioni05}. 
This argument can be easily extended to the situation 
where $p$ identical graphs $G^0$ are attached to $n=0$. 
In the limit of an infinite inserted chain, the soliton 
propagates in a star graph, which has been 
recently investigated in the context of two-dimensional networks of nonlinear 
waveguide arrays \cite{christodoulides01}. 

The general analysis carried out in \cite{burioni05} allows for an 
easy identification of the graph $G^0$ selecting 
the reflection (or the transmission) of a particular $k$ (a filter). 
For instance, a transmission filter may be obtained by 
inserting a finite chain of
3 sites ($\alpha$, $\beta$ and $\gamma$; see Fig.3).  
The momentum of perfect transmission $k^{(0)}$ is determined by 
$2 \beta_c \cos{k^{(0)}}=-\beta_{\beta,\gamma}$. 
As previously discussed, the value of the perfect transmitted
momentum does not change if $p$ identical chains are attached in $n=0$, 
but in this case the minimum becomes sharper. 
In this simple case an analytical calculation, in the linear regime, 
of ${\cal R}(k)$ gives
\begin{equation}
\frac{1}{{\cal R}}=1+ \Bigg( 
\frac{2 (\beta_{\alpha,\beta}^2 + 
\beta_{\beta,\gamma}^2-2 \beta_c^2 ) 
\sin{(2k)}-2 \beta_c^2\sin{(4k)}}
{p(\beta_{\beta,\gamma}^2-2 \beta_c^2 (1+\cos{(2k)})) } 
\Bigg)^2. 
\label{R_catena_di_3}
\end{equation}

A reflection filter can be obtained by adding a finite 
chain of $2$ sites ($\alpha$ and $\beta$).
A diagonalization of $A^0$ proves
that a total reflection is obtained at the momentum $k^{(max)}$ satisfying 
$2 \beta_c \cos{k^{(max)}}=-\beta_{\alpha,\beta}$. 
The analytical expression for ${\cal R}(k)$ is obtained from Eq.(\ref{R_catena_di_3}) 
setting $\beta_{\beta,\gamma}=0$ and $p=1$. 
In Fig.3, the good agreement between numerical findings, obtained considering the 
full non-linear evolution,
and analytical results is evidenced for large-fast solitons ($\gamma_0=40$).
 
A high-pass/low-past filter, allowing 
the transmission of high/low velocity solitons (see Fig.4),
can also be obtained. 
The graph realizing a high-pass filter 
is given by $p$ finite chains 
of length $2$ and the coupling terms   
fixed to $\beta_c$. The analytical expression for ${\cal R}(k)$ is obtained by setting 
$\beta_{\alpha,\beta}=\beta_c$ and $\beta_{\beta,\gamma}=0$ 
in Eq.(\ref{R_catena_di_3}) (the  cutoff on the momentum depends on $p$). 
The low-pass effect is obtained with
a linear chain of 3 sites ($\alpha$, $\beta$, and $\gamma$).
If $\beta_{\beta,\gamma}=2 \beta_c$, then ${\cal T}=0$ for 
$k=\pi/2$ and $\lim_{k \to \pi} {\cal R}=0$. Therefore, one has 
a low-pass with the cutoff 
momenta depending on $\beta_{\alpha,\beta}$ (in Fig.4 
$\beta_{0,\alpha}=\beta_{\alpha,\beta}=\beta_c$ and 
$\beta_{\beta,\gamma}=2 \beta_c$ ).

In conclusion, we showed that, 
by a pertinent choice of the topology of the graph $G^0$, 
one is able to control the 
reflection and the transmission of traveling kinks. 
The results obtained show the remarkable influence of topology 
on nonlinear dynamics, and apply in general to soliton propagation 
in discrete networks whose shape (i.e., topology) is controllable. 
As these results suggest, we feel that it is now both timely 
and highly desirable to develop the investigation 
of nonlinear models on general inhomogeneous networks of coupled networks, 
since one should expect new and interesting 
phenomena arising from the interplay 
between nonlinearity and topology.

{\em Acknowledgments:} 
We thank M.J. Ablowitz for stimulating discussions.

\begin{figure}[htb]
\centerline{\includegraphics[width=8.3cm]{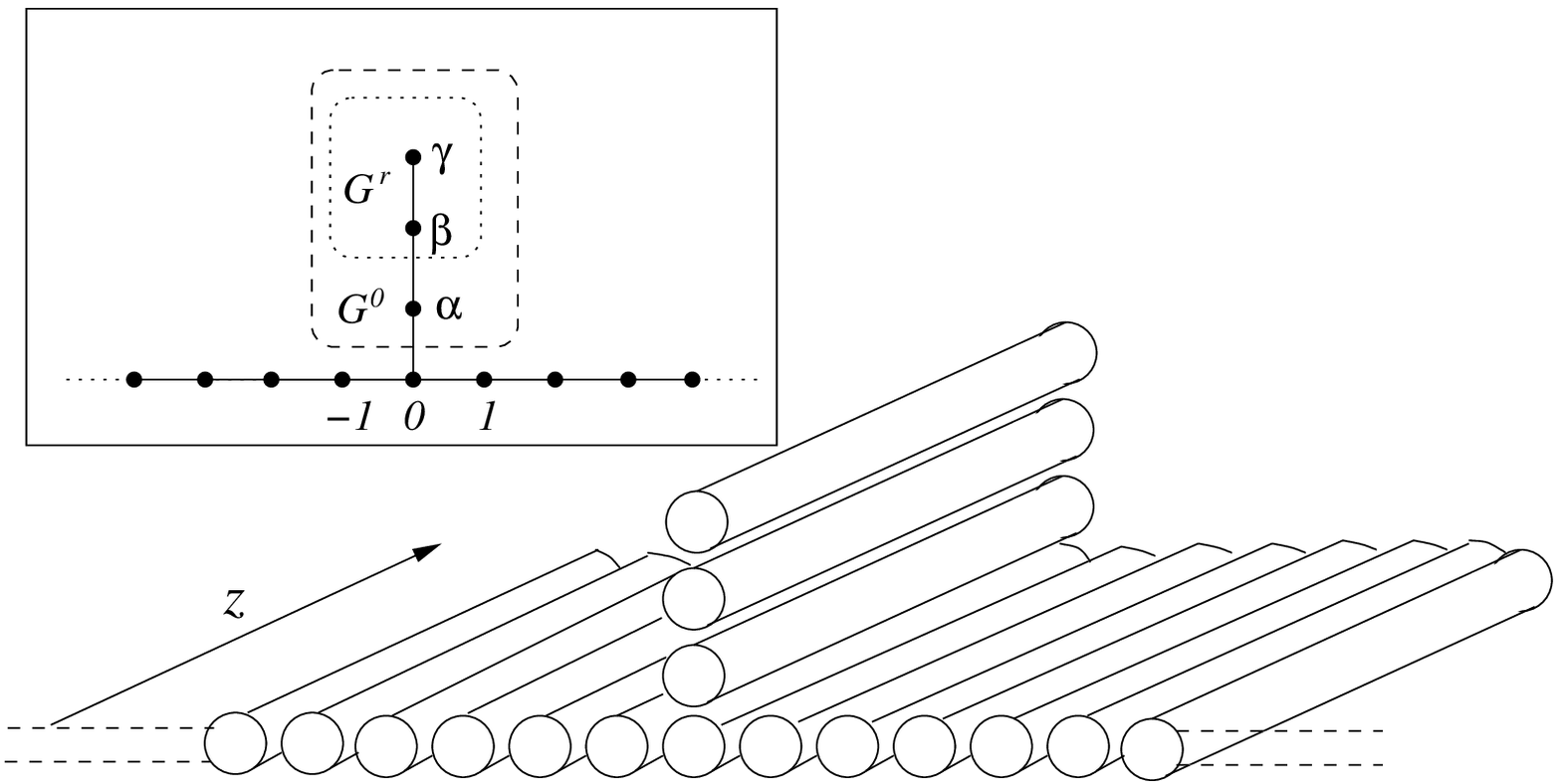}}
\caption{The inhomogeneous networks of waveguides, 
obtained attaching to a chain 
the graph $G^0$, which in this case is a finite chain of length $3$.
In the inset we plot the corresponding graph.}
\end{figure}

\begin{figure}[htb]
\centerline{\includegraphics[width=8.3cm]{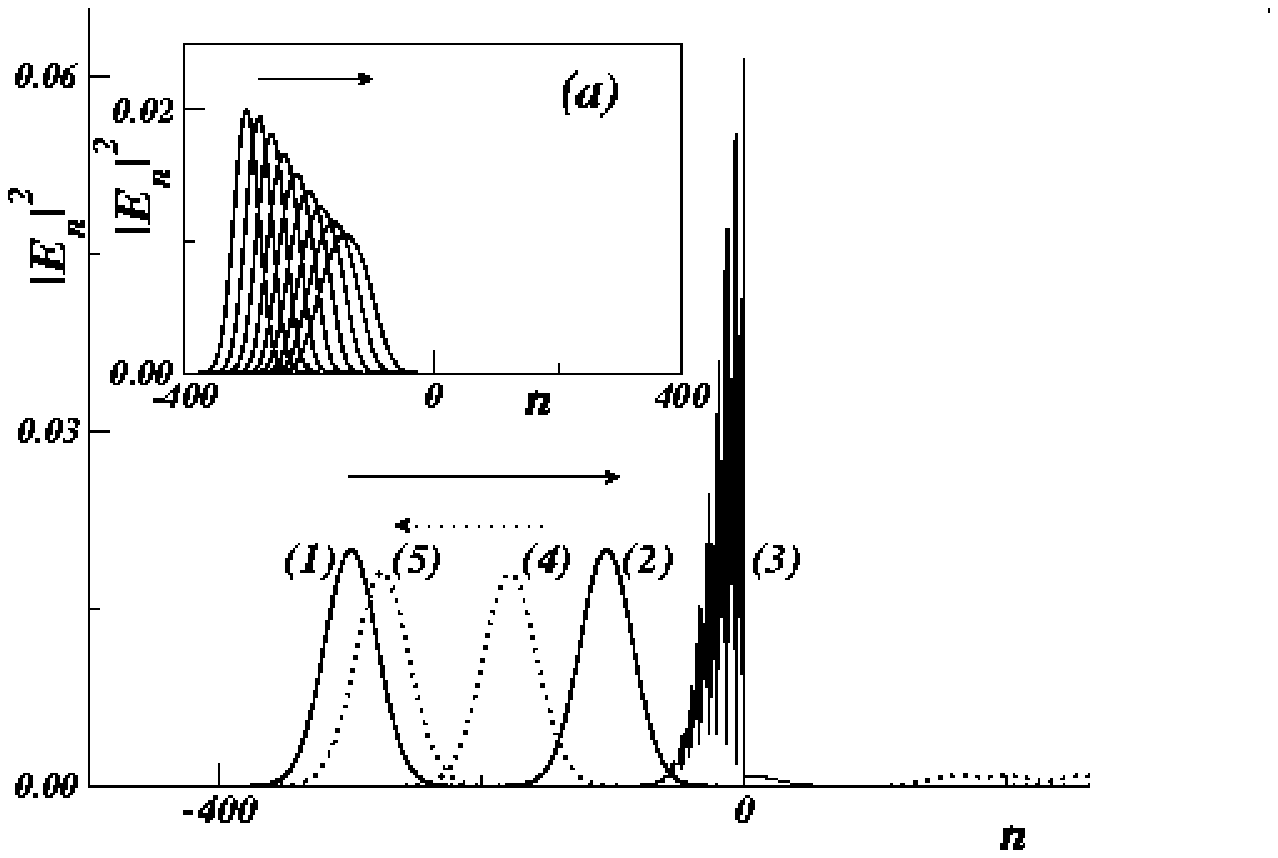}}
\caption{A soliton  
[$k_0=1.8$, $\gamma_0=40$  and $\Lambda$ given by Eq.(\ref{lambda_sol})] 
scatters through the graph of two sites described in Fig 3. The predicted 
reflection coefficient 
is close to $1$. Inset: wavepacket evolution for initial 
momentum $k_0=0.2$ (wrong non-nonlinearity), 
the wavepacket spreads before hitting $G^0$.}
\end{figure}

\begin{figure}[htb]
\centerline{\includegraphics[width=8.3cm]{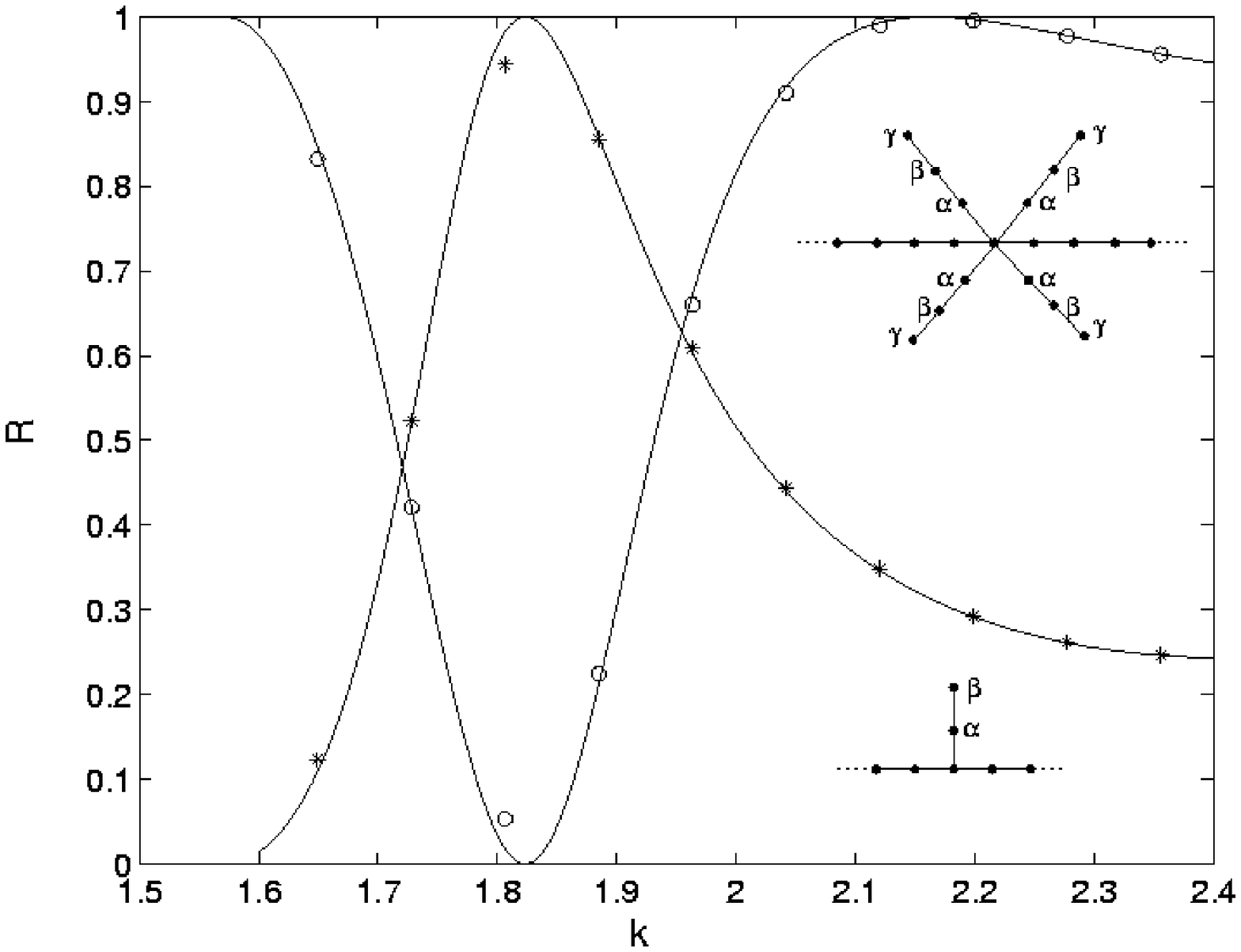}}
\caption{The reflection coefficient vs. the momentum 
$k$ for a transmission filter and a reflection filter. 
For the transmission filter   
$\beta_{0,\alpha}=\beta_{\alpha,\beta}=\beta_c$ 
and $\beta_{\beta,\gamma}=\beta_c/2$, for the reflection filter 
$\beta_{0,\alpha}=\beta_c$ and $\beta_{\alpha,\beta}=\beta_c/2$. 
Stars and circles are numerical results. 
The reflected and the transmitted momentum is $\approx 1.8$. 
Solid lines correspond to Eq. (\ref{R_catena_di_3}).}
\end{figure}

\begin{figure}[htb]
\centerline{\includegraphics[width=8.3cm]{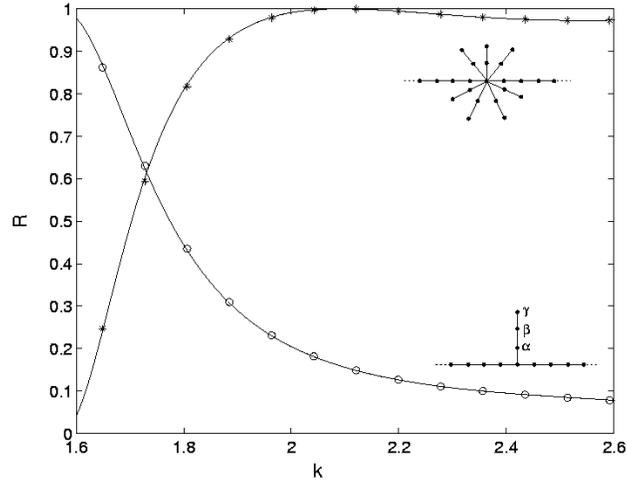}}
\caption{The reflection coefficient vs. the momentum 
$k$ for a low-pass filter and a high-pass filter (and $p=7$). 
Stars and circles are numerical results of Eq.(\ref{DNLS-gen}). 
Solid lines corresponds to Eq.(\ref{R_catena_di_3}).}
\end{figure}

\end{document}